\documentclass{article}                    
\RequirePackage{fix-cm}
\usepackage{graphicx}
\usepackage[unicode=true,pdfusetitle,bookmarks=true,bookmarksnumbered=false,bookmarksopen=false,
breaklinks=false,pdfborder={0 0 0},backref=false,colorlinks=false]{hyperref}

\usepackage{authblk}

\begin{document}

\title{Experimental Bounds on Classical Random Field Theories
}





\author[1]{Joffrey K. Peters}
\author[2]{Jingyun Fan}
\author[2]{Alan~L. Migdall}
\author[2]{Sergey V. Polyakov}
\affil[1]{JQI\\University of Maryland, College Park. joffrey@umd.edu}
\affil[2]{100 Bureau Dr. MS 8441, Gaithersburg MD, 20899}

\renewcommand\Authands{ and }

\date{\today}

\maketitle

\begin{abstract}
Alternative theories to quantum mechanics motivate important fundamental tests of our understanding and descriptions of the smallest physical systems. Here, using spontaneous parametric downconversion as a heralded single-photon source, we place experimental limits on a class of alternative theories, consisting of classical field theories which result in power-dependent normalized correlation functions. In addition, we compare our results with standard quantum mechanical interpretations of our spontaneous parametric downconversion source over an order of magnitude in intensity. Our data match the quantum mechanical expectations, and do not show a statistically significant dependence on power, limiting on quantum mechanics alternatives which require power-dependent autocorrelation functions.
\end{abstract}

\section{Introduction}
\label{intro}
Quantum mechanics (QM) gives the most accurate known description of nature at the microscopic level. However, its interpretation remains the source of much debate \cite{Bohm1952}\cite{Bohm1952a}\cite{Stapp1972}\cite{Zeilinger1999}\cite{MurrayGell-Mann1999}\cite{Schlosshauer2005}\cite{Hewitt-Horsman2009}\cite{Clarke2014}, spawning even interpretations of interpretations \cite{Beneduci2014}. Some explanations and features of the theory caused concern for a number of physicists by raising randomness to an inherent property of nature, and through questions of causality and contextuality \cite{Bohm1952}\cite{Einstein1935}\cite{Ballentine1970}. Even after decades of experimental support for QM, there is some controversy about how one should interpret a construct as central as the wavefunction \cite{Gao2011}\cite{Pusey2012}. Many foundational experiments, particularly those testing the non-locality or realism of QM, rely on the particle-like behavior of light \cite{Aspect1982}\cite{Hong1987}\cite{Gleyzes2007}\cite{Salart2008}. A large number these experiments have been done with threshold type detectors - detectors which only declare the arrival of light when the incident light packet exceeds some threshold. 

A classical theory based on a stochastic light-detector interaction has been proposed \cite{Khrennikov2012}\cite{Khrennikov2012a}\cite{Khrennikov2013} which brings the threshold detection mechanism to the fore, describing it as fundamentally responsible for observed behaviors which were deemed inherently quantum \cite{Clauser1974}\cite{Aspect1981}\cite{Aspect1982}\cite{Hong1987}. In the spirit of Lamb's anti-photon \cite{Lamb1995}, this ``prequantum" classical statistical field theory (PCSFT) avoids using the notion of discrete photons, and instead describes the light-detector interaction as a Wiener-type process \cite{Redner2001}. Clicks of the threshold detector happen as the light field crosses a threshold for the first time. Indeed, using a random-walk-like process is not surprising, as the Schr\"odinger equation is a diffusion equation, and classical theories based on diffusion have been used to reproduce aspects of QM \cite{Kershaw1964}\cite{Nelson1966}. This theory, however, places special weight on the energy threshold of the detection scheme, resulting in a normalized autocorrelation function ($g^{(2)}(0)$) which depends on the detection threshold, whereas QM predicts no such variation. Measuring the dependence of $g^{(2)}(0)$ on detection threshold would then provide a test discriminating between QM, and alternative theories.

In the single-photon avalanche photodiodes (SPADs) used in many modern low-light detection systems, the energy threshold is set by the physical properties of the semiconductors from which they are created \cite{Eisaman2011}\cite{Khrennikov2012a}. The silicon SPADs we employ have a $\mathcal{E}_{\mathrm{d}} = 1.1\:\mathrm{ eV}$ bandgap, while a typical InGaAs SPAD would have a bandgap of $0.75\:\mathrm{ eV}$ \cite{AshcroftAndMermin}. These detector types together span much of the optical spectrum, but have little overlap, such that using these types of SPADs to measure the same source is difficult. Tuning the bandgaps of the detector materials through doping is limited, brings in unwanted side effects, and is considerably more involved than turning a knob in the lab (see e.g. \cite{Wrobel1967}). To further complicate the situation, detection in SPADs depends on the current comparator threshold for discriminating avalanches caused by incident light from background circuit noise, device geometry, etc. \cite{SinglePhotonBook}. We therefore recast PCSFT in terms of incident power, since results depending on the detection threshold also scale with incident power, as we show below. 

\section{Theoretical Motivations}
\label{sec:Theory}

In PCSFT, the electric field undergoes a diffusive process in the detector before triggering a detection event at the first time it crosses a threshold. The threshold is crossed when the energy in the field is greater than the detector threshold energy $\mathcal{E}_{\mathrm{d}}$. In a bounded 1-dimensional Wiener process, such as described by PCSFT, the mean time to hit the threshold $\mathcal{E}_{\mathrm{d}}$ scales as $\mathcal{E}_{\mathrm{d}}^2$ \cite{Redner2001}. Since the energy in the electric field is proportional to the square of the field, the mean time to first cross this threshold is
\begin{equation}
\label{eq:meanHittingTime}
\bar{\tau} = \frac{\mathcal{E}_{\mathrm{d}}}{\sigma^2},
\end{equation}
where $\sigma^2$ is the power in the light field \cite{Khrennikov2012}. Since the predicted detection rates depend on this mean detection time $\bar{\tau}$, it is suggested that the experimenter change $\mathcal{E}_{\mathrm{d}}$ to vary the click rates and test the theory. We propose instead to change the incident power to vary $\bar{\tau}$. These parameters have an inverse effect on the click times, and this should be reflected in the results of the theory.

A version of PCSFT was developed using density matrix formalism in analogy to QM to give normalized autocorrelation functions $g^{(2)}(0)$ within the theory's framework. Particularly, the theory places a bound on $g^{(2)}(0)$ in Inequality 36 of Ref. \cite{Khrennikov2012}:
\begin{equation} \label{ineq:36}
g^{(2)}_{\mathcal{E}_{\mathrm{d}}}(0)\leq(\frac{2\delta}{\mathrm{\Delta} t})\frac{\bar{\mathcal{E}}_{\mathrm{pulse}}}{\mathcal{E}_{\mathrm{d}}},
\end{equation}
where $\delta$ is the temporal duration of the light pulse, $\mathrm{\Delta} t$ is the detection time bin, and $\bar{\mathcal{E}}_{\mathrm{pulse}}$ is the energy in a light pulse.

We see that Ineq. \ref{ineq:36} is inversely proportional to $\mathcal{E}_{\mathrm{d}}$. From Eq. \ref{eq:meanHittingTime}, we would expect that changing the incident light power should have the inverse effect on mean detection times as changing $\mathcal{E}_{\mathrm{d}}$, and thus this bound on $g^{(2)}(0)$ is proportional to the incident power. Indeed, using equations from Ref. \cite{Khrennikov2012}, we can recast Ineq. \ref{ineq:36} in terms of power. Using Eq. 34 from Ref. \cite{Khrennikov2012}, noting that total power is the sum of the powers in each channel, using Eqs. 14, 15 from that reference, and noting that two channels ($j=1,2$) must enter into a $g^{(2)}(0)$ measurement, one obtains the inequality:
\begin{equation}
\label{ineq:SergeyIneq}
g^{(2)}_{\mathcal{E}_{\mathrm{d}}}\leq\frac{2\delta^2}{\mathrm{\Delta} t}\frac{N_1 + N_2}{T}
\end{equation}
where $N_{1,2}$ are counts from two detectors each monitoring half the incident light energy, $T$ is the total integrated data-taking time, and equal detection efficiencies in channel 1 and 2 are assumed for simplicity. In both a QM approach, and in PCSFT, the light power is proportional to the number of counts $N_{1,2}$ in a time window. However, from a quantum mechanical perspective, we do not anticipate $g^{(2)}(0)$ to scale with the incident energy, while from Ineq. \ref{ineq:SergeyIneq} we see that as formulated, in PCSFT the correlation function may grow with increasing incident energy. Thus, a measurement of $g^{(2)}(0)$ over a range of intensities serves as a fundamental test of QM. Agreement with the QM theory would place experimental bounds on any detector threshold based models such as Ref. \cite{Khrennikov2012}.

In this work, we measure the power dependence of the second order autocorrelation function, $g^{(2)}(0)=A(N_1/\eta_1+N_2/\eta_1)+B$, where $\eta_{1,2}$ are total channel efficiencies of beam paths to detectors 1 and 2 (including beam path, transmittances, collection and detector efficiencies, but excluding an adjustable attenuation). We place limits on the slope $A$ of this dependence. In Sec. \ref{methods} below, we describe our apparatus and data-taking methods to measure $g^{(2)}(0)$ as a function of incident light power.

\section{Methods}
\label{methods}

\begin{figure}
\includegraphics[width=1\textwidth]{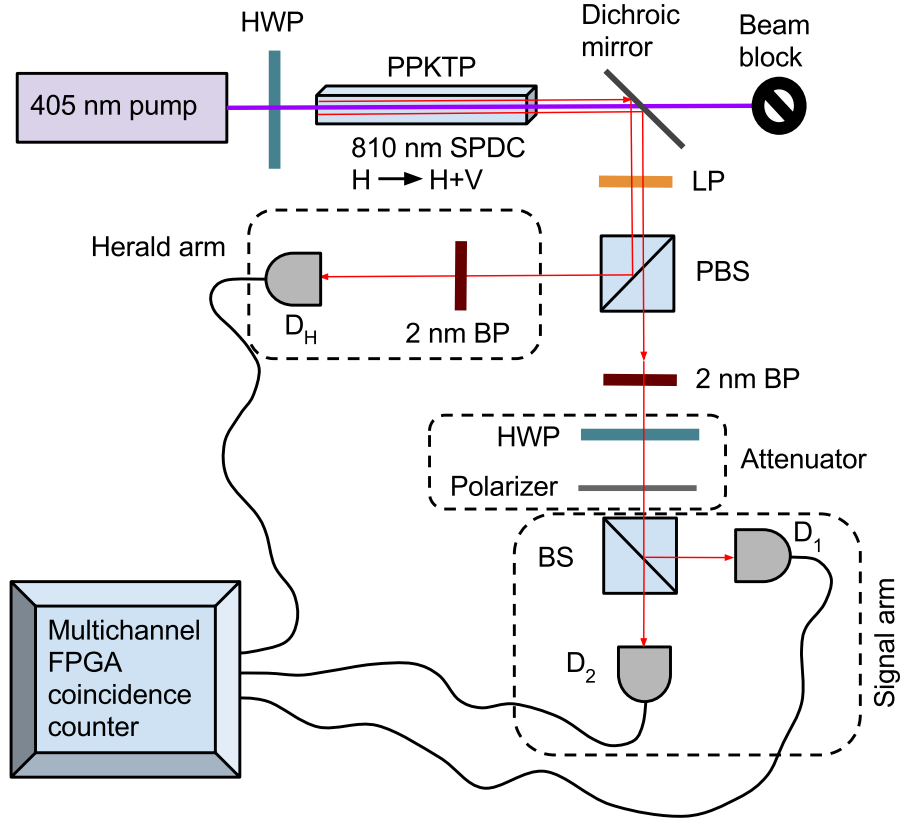}
\caption{Experimental setup. A $405\:\mathrm{ nm}$ laser pumps a PPKTP crystal with horizontally polarized light to generate correlated horizontal (H) and vertical (V) beams at $810\:\mathrm{ nm}$. The $810\:\mathrm{ nm}$ light is picked off with a dichroic mirror, and is passed through a long-pass (LP) filter, then the H and V  light is separated on a polarizing beam splitter (PBS). The output modes of the PBS pass through $2\:\mathrm{ nm}$ bandpass (BP) filters, and the heralding arm is detected on detector $D_\mathrm{H}$. The signal beam passes through a variable attenuator before being split on a 50-50 beamsplitter (BS), with output modes detected by detectors $D_1$ and $D_2$. Clicks from the detectors are recorded and coincidences processed and reported by a custom-firmware field-programmable gate array (FPGA) chip.}
\label{fig:exptSetup}
\end{figure} 

We use a spontaneous parametric downconversion (SPDC) system consisting of a $25\:\mathrm{ mm}$ periodically polled potassium titanyl phosphate (PPKTP) crystal pumped by a $405\:\mathrm{ nm}$ continuous wave diode laser. The pump has horizontal polarization, and the crystal generates type II co-linear spontaneous downconverted light. The crystal temperature is stabilized and the pump polarization set to horizontal with a half-wave plate (HWP) to generate downconversion at $810\:\mathrm{ nm}$ in two modes with horizontal and vertical orientations. After the crystal, the $405\:\mathrm{ nm}$ pump light is suppressed from the experiment by $>\!150\:\mathrm{dB}$ \cite{Pereira2013}. The two downconverted beams are separated on a PBS as shown in Fig. \ref{fig:exptSetup}. One output is coupled through single mode fiber to a Si SPAD which heralds the creation of downconverted light ($D_\mathrm{H}$ in Fig. \ref{fig:exptSetup}). The other output passes through a HWP on a digitally controllable rotation stage and a linear polarizer to set the attenuation of light in the signal arm. The output from this controllable attenuator is split on a 50-50 beam splitter and the output modes are incident on SPADs $D_1$ and $D_2$. The output of each SPAD is sent through equal length cables to a multichannel FPGA-based coincidence detector \cite{StatsFPGA2014}. Thus, our source produces heralded light pulses which are attenuated before entering a two-detector tree to measure coincidences.

All clicks and coincidences are recorded on a computer. Once sufficient data are recorded at a given signal attenuation, the HWP in the signal arm of Fig. \ref{fig:exptSetup} is rotated to a new attenuation level, and the experiment is repeated. We vary the power to the signal arm by one order of magnitude. For each of these attenuation levels, we collect up to $10^4$ coincidences between detectors $D_2$ and $D_3$ in the signal arm. For the lower light levels, data may be collected for many days. The data are recorded in $20.83\:\mathrm{ ns}$ time bins and are accumulated in the FPGA and reported  to a computer roughly every $1\:\mathrm{ ms}$. By calculating $g^{(2)}(0)$ for segments of the data set at a time, we remove bias from slow fluctuations, particularly important for the longer runs.

SPDC sources produce correlated light fields in two modes. We use one of these modes to herald light in the other mode. This type of source is often described as a ``conditional single-photon source," as a photodetection event in the heralding arm heralds with high probability the occurrence of a single event in the signal arm because our source is operated in a regime where the probability of generating any downconverted light in a given time bin is low. For our test, we use this source in a conditional mode: i.e. we find the second-order normalized autocorrelation function of the signal arm, conditioned on a detection in the heralding arm. The conditional autocorrelation function now takes the place of the unheralded $g^{(2)}(0)$ in PQSFT. Additionally, we use the correlated light from our source to calibrate the detection arm efficiencies from the source to each detector using the correlated-photon method, see Ref. \cite{Polyakov2007} and references therein. The efficiencies of the detection arms are $\eta_\mathrm{H} = 0.26(2)$, $\eta_1 = 0.075(8)$, and $\eta_2 = 0.055(6)$ (statistical uncertainties) for detector arms $D_\mathrm{H}$, $D_1$, and $D_2$ respectively.

To find background levels, we rotate the pump beam polarization by $90^{\circ}$ to turn off the downconversion process, then the detectors monitor detection events and coincidences with no attenuation. This is compared with the pure dark counts, as measured with the pump laser completely off. The two differ in that the background includes incoherent spontaneous emission scattered into the detected modes. Without any signal arm attenuation, this spontaneous emission background contributes noise on the order of the dark counts in the detectors (which were between $114\:\mathrm{ s}^{-1}$ and $183\:\mathrm{ s}^{-1}$).

The low background, $\approx50\:\%$ detector efficiency, $30\:\%$ source extraction efficiency, and heralded $g^{(2)}(0)$ comparable to the best sources, give us a competitive experimental platform (c.f comparable sources and detectors in Ref. \cite{Eisaman2011}) with which to look for power dependence of the autocorrelation function.

\section{Results}
\label{results}
\begin{figure}
\includegraphics[width=1\textwidth]{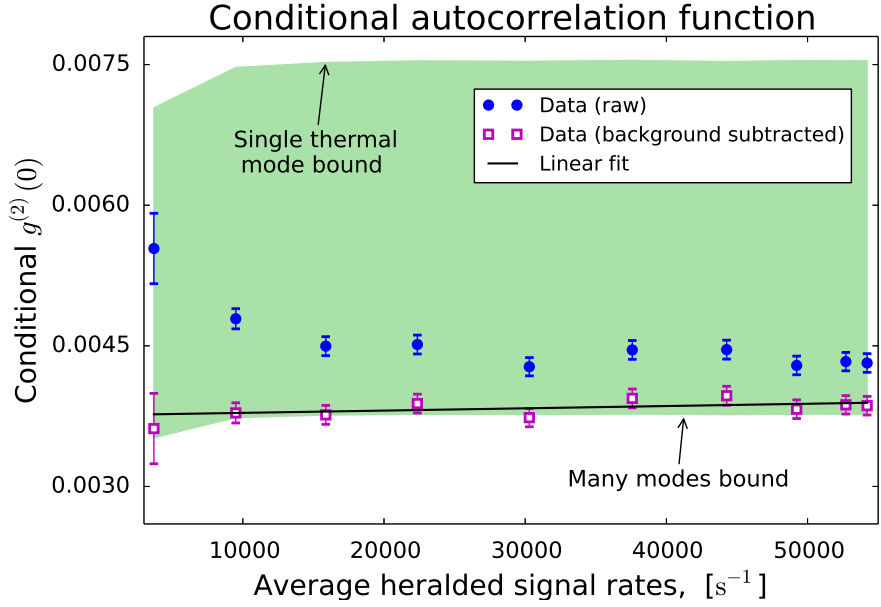}
\caption{Autocorrelation function conditioned on detection in the heralding arm as a function of average heralded signal arm count rate, corrected for detector arm efficiencies. Error bars represent $1$ standard deviation statistical uncertainties from number of counts. Disks represent raw data, while open squares are data with background subtracted. The shaded region represents the quantum mechanical expectation. It was generated by back-propagating losses to estimate the rate of double-pairs from the pair production rate, then using the double-pair rate to find $g^{(2)}(0)$. Depending on the mode structure of the detected light, QM predicts a $g^{(2)}(0)$ in the shaded region. The upper bound corresponds to a single matched thermal mode, while the lower bound corresponds to many detected modes (thermal and Poisson possible).}
\label{fig:condG2}
\end{figure}

For each attenuation, we calculate the autocorrelation conditioned on a click in the heralding detector $D_\mathrm{H}$. Results are presented in Fig. \ref{fig:condG2}. We compare these results with the expectations from a quantum mechanical model. The rate on the horizontal axis was calculated based on the average detected rate of counts in the signal arm, accounting for losses to give the number of photons incident on the signal arm.

This source is conditional: a click in the heralding detector indicates the presence of light incident on the signaling arm. Due to the probabilistic nature of the source and the lack of number-resolution of the detectors, there is a non-zero probability of generating two or more pairs of photons, resulting in a low, but finite residual $g^{(2)}(0)$. (Accidental coincidences from dark counts and background are negligible.) Accounting for the efficiencies in the heralding and signal arms, and taking only lowest order terms, we find that $g^{(2)}(0)$ conditioned on a heralding click is described approximately by
\begin{equation}
\label{eq:g2QM}
g^{(2)}(0) \approx 2 \frac{P(2)}{P(1)}\frac{(1 -(1- \eta_\mathrm{H})^2)}{\eta_\mathrm{H}}
\end{equation}
where $P(1)$ and $P(2)$ are the probabilities of the SPDC source creating a single pair of photons, and a double pair respectively, $\eta_\mathrm{H}$ is the overall detection efficiency of the heralding arm, and we have ignored small corrections from noise in the signal arm \cite{Goldschmidt2014}. The efficiency term in the numerator is greater than $\eta_\mathrm{H}$ because the detector has two chances to detect one from the incident double. Propagating losses backward from the detector to the source, we estimate $P(1)$, the rate of single light pulses in the signal arm. The probability of a double pair is approximately
\begin{equation}
\label{eq:P2}
P(2) \approx \frac{G}{2} P(1)^2
\end{equation}
where $1\leq G\leq2$ depends on the mode structure of the source \cite{Goldschmidt2013a}.

The output of SPDC is typically described by perfectly correlated thermal modes. For a single pair of perfectly matched thermal modes, we would expect $G=2$. As the number of modes increases, $G$ asymptotically approaches $1$. Plugging these values for $G$ into Eq. \ref{eq:g2QM} defines the bounds of the shaded region of Fig. \ref{fig:condG2}. It is important to note that the shaded region does not imply that $g^{2}(0)$ may vary randomly within this range, but rather that it should parallel the edge contours of the shaded region (which would ideally be flat, but are based on measurment here).

Given the $91\:\mathrm{GHz}$ bandwidth of the light transmitted through the $2\:\mathrm{nm}$ bandpass filters in our setup, and the $<1\:\mathrm{GHz}$ bandwidth of the continuous wave pump, we expect there to be a large number of modes collected, driving $G$ toward $1$. In Fig. \ref{fig:condG2}, we see that the measured heralded $g^{(2)}(0)$ values lie close to this $G=1$ limit, and within their uncertainty of the QM bounds.

To test the validity of QM and limit alternatives, we check for dependence of $g^{(2)}(0)$ on power. The solid line in Fig. \ref{fig:condG2} shows a linear fit to the background-subtracted data. The reduced $\chi ^2$ of the fit is $0.8$, indicating a good fit to the data. The fit slope is $2.4(1.5)\times10^{-9}\:\mathrm{s}$. Thus we observe no power dependence of $g^{(2)}(0)$. The constant (offset) of the fit is $g^{(2)}(0) = 0.00376(6)$, which further bounds the parameters of Ineq. \ref{ineq:36}. It is worth noting that as with any realistic source, we have a low but nonzero $g^{(2)}(0)$, comparable with the state of the art sources (c.f. Table 1 of Ref. \cite{Eisaman2011}). We stress that the residual $g^{(2)}(0)$ measured is fully described by generation of double pairs and is compatible with a QM description of the source. To completely rule out PCSFT as it is currently described would require a source with $g^{(2)}(0)=0$, however a more stringent limit on PCSFT may be set, with a more pure single-photon source. Perhaps using a high purity, low-noise source such as that in Ref. \cite{Brida2012} would be a good next step to further constrain alternative theories. Additionally, we invite PCSFT proponents to refine the theory to give the explicit dependence of $g^{(2)}(0)$ on power.

\section{Conclusion}
We searched for a power-dependence in the normalized autocorrelation of a beam of ``single-photon" light. The quantum description is compatible with the observed $g^{(2)}(0)$, and the data show no dependence on power over more than one order of magnitude variation of power. The data were fitted to a line, the slope of which sets an upper bound on the possible dependence of $g^{(2)}(0)$ on power.  The slope of this dependence is consistent with zero to within $2$ standard deviations, limiting PCSFT or other classical theories which result in non-zero, power-dependent autocorrelations.

\section*{Acknowledgements}
The authors thank Andrei Khrennikov for fruitful and informative conversations.


\bibliographystyle{spphys}       
\bibliography{AMOPapers.bib}   

\end{document}